# The extraordinary radiation pattern of an optical rod antenna

*Chenglong Zhao\* and Jiasen Zhang\**

Department of Physics and State Key Laboratory for Mesoscopic Physics, Peking University, Beijing 100871, China

We investigated the radiation pattern of an optical rod antenna and found that it had many features compared with its conventional radio-wave equivalents. After defining a parameter $\Lambda = \lambda_{\text{eff}}/\lambda$, which was the ratio of the effective wavelength of the rod antenna to the incident wavelength, we found that $\Lambda$ had a great influence on the radiation pattern. Even the radiation pattern with a higher resonant order is without side lobes and results in a sharper directivity, which provides new design flexibilities in improving the directivities of the optical antennas.

\*To whom correspondence should be addressed e-mail:
 chenglong@pku.edu.cn; jszhang@pku.edu.cn.

Radio-wave antennas, which play an important role in modern society, have been investigated for decades. However, only recently have research shown that it is possible to design antennas in optical region based on the plasmonic resonance of metallic materials thanks to the development of nanofabrication. Optical half-wave, bow-tie and quarter wavelength antennas have been experimentally realized [1-3]. Metal can be treated as perfect conductors in radio-wave frequency; therefore, the wavelength in the radio-wave antenna is almost equal to the incident wavelength. For a conventional ideal half-wave antenna, its resonant condition achieves when its length $L$ is equal to half of its radiation wavelength $\lambda$, i.e. $L = \lambda/2$; however, in the optical frequency, metal behaves completely different with plasmonic resonance. As a result, the wavelength in an optical antenna is greatly shorter than the incident wavelength. For instance, the length of a resonant dipole optical antenna is shorter than one half of the incident wavelength [4-6]; therefore, optical antennas are quite different from its conventional antenna equivalents [7]. Lukas Novotny [8] has show that the incident wavelength $\lambda$ can be replaced by an effective wavelength $\lambda_{eff}$ in order to directly apply the radio-wave antenna theory to the optical counterparts. Applying the ideas from the radio-wave antennas to that of optical region, optical antennas can be designed to enhance electric field, influence the single-molecule fluorescence [9-11] and modify radiation directivity [12]. However, only a few groups have concerned the radiation pattern of the optical antennas [13, 14].

In this letter we investigate the radiation pattern of a single Au rod which can be considered as an impedance matching antenna compared with two aligned and closely spaced rods [8], and we compared its radiation pattern with that of the conventional radio-wave wire antenna because of their similarities in conception. After defining a parameter $\varLambda = \lambda_{eff}/\lambda$, which is the ratio of the effective wavelength of the rod to the incident wavelength, we found that the radiation pattern of the Au rod antenna has some new features compared with that of a conventional radio-wave wire antenna, where $\varLambda = 1$. Unlike in the case of a conventional radio-wave wire antenna, where the first order resonance is mostly used mainly because of higher order resonance resulting in large side lobes, however, higher order resonance can be used in an optical rod antenna to achieve better directivity no need worrying about side lobes. The extraordinary radiation pattern of an optical antenna has potential applications in improving the directivities of the optical antennas.

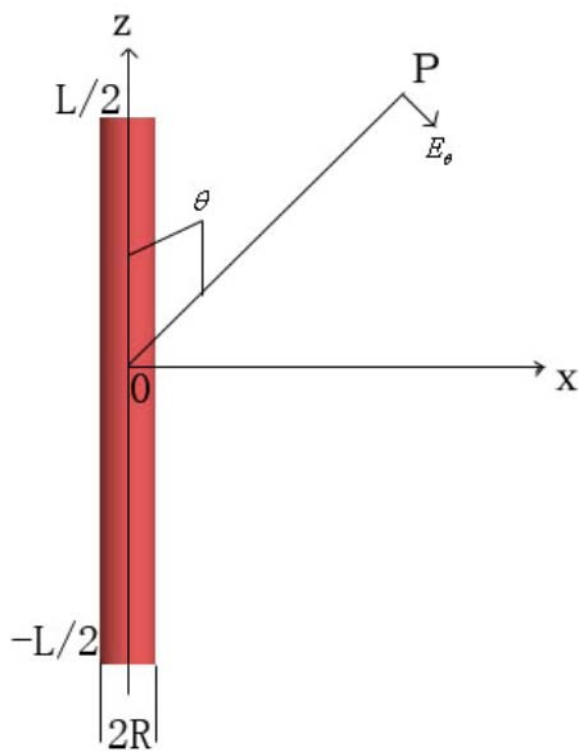

Figure 1. An optical rod antenna with total length $L$ and radius $R$. $P$ is a point in the far zone, $E_\theta$ is the $\theta$ component of the electric far field in spherical coordinates.

A single Au rod with dielectric function $\varepsilon(\omega)$, radius $R$, and total length $L$ is shown in Figure 1. The rod is embedded in a medium with dielectric constant $\varepsilon_s$ and illuminated along $x$ axis by a plane wave with wavelength $\lambda$ and polarization in $z$ direction. The incident plane wave excites surface plasmon polariton propagating along the rod and forms a standing wave [15]. The induced currents in the rod radiate as that of a conventional radio-wave wire antenna does. In order to determine the radiation pattern, the induced current in the rod must be known. If a thin rod is assumed, i.e. $R \ll L$, then the electric field of the $TM_0$ mode inside the rod is approximately:

$$E = E_1 J_0(\kappa_1 r)\sin[\gamma(\frac{L+2R}{2}+|z|)] \quad, \quad r<R \tag{1}$$

where $E_1$ is a constant determined by the incident plane wave amplitude and boundary conditions, $J_0(\kappa_1 r)$ is the cylindrical Bessel function of zero order, $\gamma$ is the propagation constant of the surface plasmon polariton along the rod and $\kappa_1=(k_0^2\varepsilon(\omega)-\gamma^2)^{1/2}$.

According to the waveguide theory, $\gamma$ can be determined by the following characteristic equation:

$$\frac{\varepsilon(\omega)}{\kappa_1 R}\frac{J_1(\kappa_1 R)}{J_0(\kappa_1 R)} - \frac{\varepsilon_s}{\kappa_2 R}\frac{H_1^{(1)}(\kappa_2 R)}{H_0^{(1)}(\kappa_2 R)} = 0 \quad, \tag{2}$$

Where $\kappa_2=(k_0^2\varepsilon_s-\gamma_2)^{1/2}$, $H^{(1)}(\kappa_2 R)$ is the cylindrical Henkel function.

The induced current density $j(r, z)$ inside the rod is $j(r, z) = -i\omega\varepsilon_0[\varepsilon(\omega)-1]E$, then the total current along the rod is:

$$I(z) = \int_0^R \int_0^{2\pi} j(r,z) r dr d\varphi = I_m \sin[\gamma(\frac{L+2R}{2}+|z|)] \quad, \tag{3}$$

where $I_m = -i\omega\varepsilon_0[\varepsilon(\omega)-1]E_1 \times 2\pi R J_1(\kappa_1 R)/\kappa_1$ is a constant for a given $\omega$ and $R$.

After the current is determined, the radiation pattern can be calculated following the procedure described in Ref. 16. The far field $E_\theta$ can be calculated as

$$E_\theta = j\omega\mu\frac{\exp(-j\beta r)}{4\pi r}\sin\theta \times \int_{-\frac{L}{2}}^{\frac{L}{2}} I(z)\exp[j\beta z\cos(\theta)] \tag{4}$$

where $\beta=2\pi(\varepsilon_s)^{1/2}/\lambda$ is the propagation constant in the medium. Then we obtain

$$E_\theta = j\omega\mu\frac{\exp(-j\beta r)}{4\pi r}\frac{I_m}{\beta}\times\left\{\frac{\beta\sin\theta}{\gamma+\beta\cos\theta}[\cos(\frac{\beta L\cos\theta}{2}-\gamma R)-\cos(\frac{n\pi}{2})]+\frac{\beta\sin\theta}{\gamma-\beta\cos\theta}[\cos(\frac{\beta L\cos\theta}{2}+\gamma R)-\cos(\frac{n\pi}{2})]\right\} \tag{5}$$

Since only the far field patterns are interested, we can only concern the following function of the radiation pattern

$$F(\theta)=\frac{\beta\sin\theta}{\gamma+\beta\cos\theta}[\cos(\frac{\beta L\cos\theta}{2}-\gamma R)-\cos(\frac{n\pi}{2})]+\frac{\beta\sin\theta}{\gamma-\beta\cos\theta}[\cos(\frac{\beta L\cos\theta}{2}+\gamma R)-\cos(\frac{n\pi}{2})] \tag{6}$$

In the resonant situation, which means $L = n\lambda_{eff}/2$, where $n$ is the resonant order, i.e. $n = 1$ corresponding to a half-wave antenna, $\lambda_{eff}$ is the effective wavelength of the rod [8] and

$$\lambda_{eff} = \frac{2\pi}{\gamma} - \frac{4R}{n} \qquad (7)$$

If we define $D = \beta/\gamma$, which is the velocity factor in antenna theory; $\Lambda = \lambda_{eff}/\lambda$, which is the ratio of the effective wavelength of the rod to the incident wavelength. Then $D = \Lambda + 4R/n\lambda$ and

$$F(\theta) = \frac{D\sin\theta}{1+D\cos\theta}[\cos(\frac{n\pi}{2}\Lambda\cos\theta - \gamma R) - \cos(\frac{n\pi}{2})] + \frac{D\sin\theta}{1-D\cos\theta}[\cos(\frac{n\pi}{2}\Lambda\cos\theta + \gamma R) - \cos(\frac{n\pi}{2})] \quad (8)$$

Since it is impossible to excite even order resonance when the incident plane wave polarizes along the rod [15], we mainly concern the odd order resonance, i.e. $L = n\lambda_{eff}/2$ and $n = 1, 3, 5 \ldots$. Then equation (8) becomes:

$$F(\theta) = \frac{D\sin\theta}{1+D\sin\theta}\cos(\frac{n\pi}{2}\Lambda\cos\theta - \gamma R) + \frac{D\sin\theta}{1-D\sin\theta}\cos(\frac{n\pi}{2}\Lambda\cos\theta + \gamma R) \quad (9)$$

Now we consider a real situation: an Au rod with $R = 5$ nm embedded in vacuum, the wavelength of the incident plane wave is $\lambda = 850$ nm. Then using equation (2) and (7) we can calculate the following parameters for different resonant order $n$ in Table 1:

Table 1. The parameters for different resonant order n

| $n$ | 1 | 3 | 5 | 7 | 9 |
|---|---|---|---|---|---|
| $L$ (nm) | 63 | 207 | 352 | 497 | 642 |
| $\lambda_{eff}$ (nm) | 125 | 138 | 141 | 142 | 143 |
| $\Lambda$ | 0.147 | 0.162 | 0.166 | 0.167 | 0.168 |

The above table shows that the effective wavelength of an optical rod antenna is greatly shorter than the incident wavelength. As we all known, the wavelength in a traditional wire antenna is almost equal to its radiation wavelength, and its radiation pattern function is [16]:

$$F(\theta) = \frac{\cos(\frac{n\pi\cos\theta}{2}) - \cos(\frac{n\pi}{2})}{\sin\theta} \qquad (10)$$

The normalized radiation patterns of the rod antenna and its radio-wave counterpart for different resonant order $n$ is shown in Figure 2 and Figure 3, respectively.

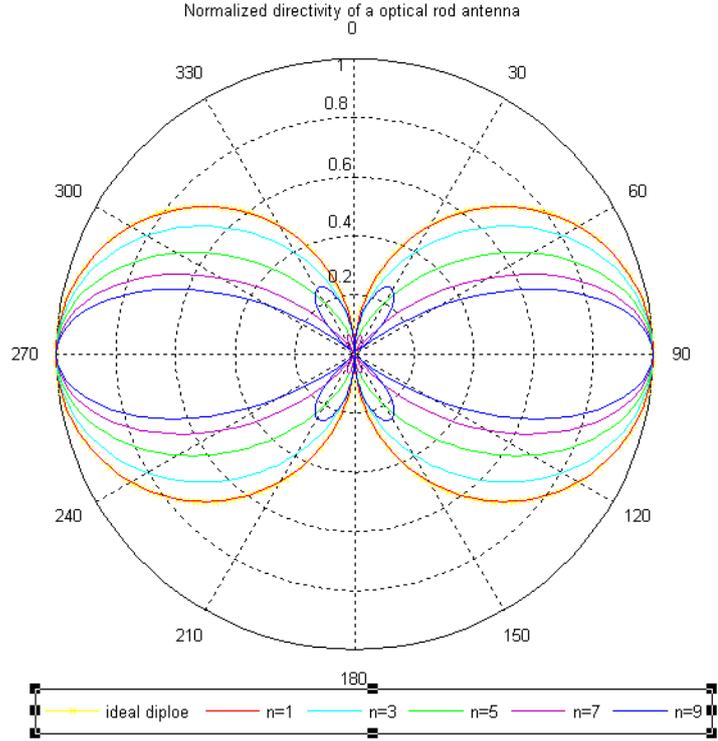

Figure 2. The directivity of an Au rod antenna for different resonant order $n$ from 1 to 9, the yellow line is the directivity of an ideal dipole with its directivity function $F(\theta) = sin(\theta)$, which is superposed with that of the $n = 1$ curve.

In figure 2, it can be seen that the $n = 1$ curve, which represents a half-wave optical rod antenna, is superposed with an ideal dipole curve. And until to $n = 7$ does it appear side lobes, which is completely different from the radiation pattern of a conventional wire antennas in figure 3. Only the case of $n = 1$ and 3 are plotted in figure 3 for comparison, it can be seen that a conventional radio-wave wire antenna has relatively large side lobes in the same resonant order compared with an optical rod antenna.

For a conventional wire antenna, $n = 3$ corresponds to a $3\lambda/2$ antenna and the multiple lobe structure is due to the canceling effect of oppositely directed currents on the antenna [16]. In the resonant condition $n = 3$, although the optical rod antenna can be treated as a $3\lambda_{eff}/2$ antenna, which also has opposite currents floating along the rod, its length is greatly shorter than the incident wavelength. As a result, the radiation patterns of the $3\lambda_{eff}/2$ and $5\lambda_{eff}/2$ rod antenna still don't have side lobes. The rod antenna is electrically small in terms of antenna nomenclature, account for its relatively short length compared with its radiative wavelength. This can be seen from the fact that radiation pattern of the half-wave rod antenna ($n = 1$) is directly superposed with that of an ideal dipole.

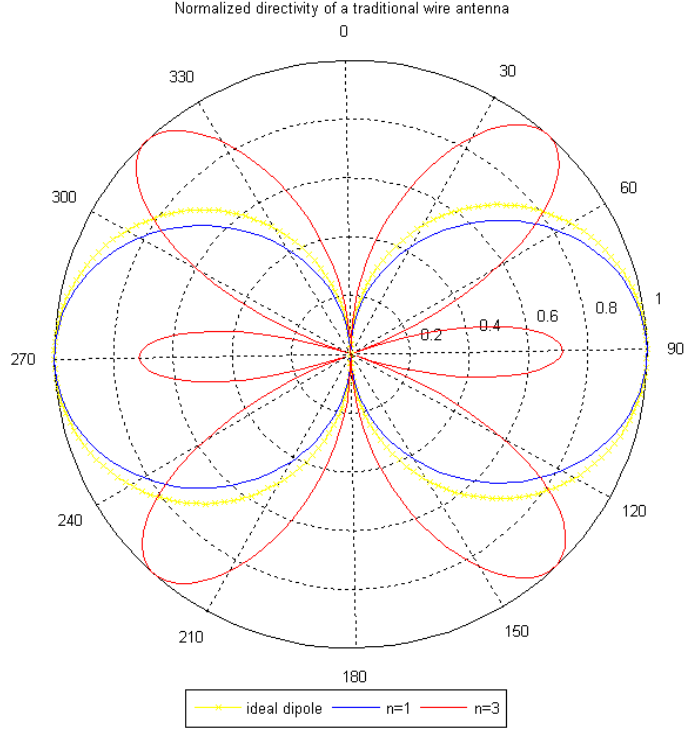

Figure 3. The directivity of an ideal radio-wave thin wire antenna with $n = 1$ and 3 according to equation (10), the yellow line is the directivity of an ideal dipole as that of in Figure 2.

This extraordinary radiation pattern of an optical rod antenna can be understood by comparing equation (9) with equation (10). In the case of a conventional radio-wave antenna, the zero point of the radiation pattern is realized when $cos[n\pi cos(\theta)/2] = 0$ from equation (10), i.e. $cos\theta = 1/n$. For resonant conditions n > 1, there always exists an angle $\theta$ corresponding to the zero point of the radiation pattern, this is where the side lobe come from when $n = 3$. In the case of the optical rod antenna, $\gamma R$ has a neglect influence on the radiation pattern; therefore, the zero point of equation (9) happens approximately when $cos[n\pi \varLambda cos(\theta)/2] = 0$, i.e. $cos\theta = 1/n\varLambda$. It can be seen that for different resonant order $n = 1$, 3 and 5 from table 1, $1/n\varLambda = 6.8$, 2.1 and 1.2, respectively, which means that for all real $\theta$, $cos\theta = 1/n\varLambda$ can never be fulfilled; thereby, the radiation pattern of those resonant order don't have side lobes. For the resonant ordres $n = 7$ and 9, $1/n\varLambda = 0.8$ and 0.7, respectively. Equation $cos\theta = 1/n\varLambda$ can be fulfilled at some $\theta$ angles. This is why the side lobes appear until n=7 in the radiation pattern. Here we show the great influence of $\varLambda$ on the radiation pattern of the optical rod antenna compared with the conventional radio-wave antenna, where $\varLambda = 1$.

It can be seen that the radiation pattern becomes sharper as *n* increases from figure 2. Unlike in a conventional radio-wave case, where only the half-wave (*n* = 1) antenna is used mainly because of the large side lobes in higher resonant order *n*, in the optical rod antenna case, higher order resonance can be used to get better directivity no need to worry about side lobes, besides, higher order resonance means longer rod which is easier to fabricated.

In conclusion, we have deduced the radiation pattern function of an optical rod antenna and found it has many different characteristics compared with the conventional radio-wave counterparts: (1) The directivity of the optical half-wave antenna is almost the same as that of an ideal dipole, which means it is still electrically small in terms of antenna nomenclature. (2) The parameter $\varLambda$ has a great influence on the radiation pattern; in an optical rod antenna, higher order resonances still don't have side lobes in the radiation pattern. (3) As the resonant order *n* increases the radiation pattern becomes sharper. Those different characteristics provide a guideline in improving the directivities of the optical antennas.